# Study of Turbulence and Pressure Recovery in the Heat Pipe Vapor Flow Using the Spectral-Element Method

**Carolina Bourdot Dutra, Tri Nguyen and Elia Merzari**
Ken and Mary Alice Lindquist Department of Nuclear Engineering
The Pennsylvania State University
311 Hallowell Bldg, University Park, PA 16802
cdutra@psu.edu; nguyen.tri@psu.edu; ebm5351@psu.edu

**Joshua E. Hansel**
Idaho National Laboratory
955 MK Simpson Boulevard, Idaho Falls, ID 83415
joshua.hansel@inl.gov

## ABSTRACT

Heat pipes can efficiently and passively remove heat in nuclear microreactors. Nevertheless, the flow dynamics within heat pipes present a significant challenge in designing and optimizing them for nuclear energy applications. This work aims to explore the vapor core of heat pipes through comprehensive two- and three-dimensional simulations, with a primary focus on modeling the pressure recovery observed in the condenser section. The goal is to establish improved correlations for one-dimensional heat pipe codes.

The simulations are validated against experimental data from a vapor pipe documented in the literature. The $k$–$\tau$ turbulence model is employed in the two-dimensional simulations through the open-source spectral-element code Nek5000. This model provides insights into pressure recovery within heat pipes with low computational cost. In addition, Large Eddy Simulations (LES) are used to capture turbulent flow features in a three-dimensional vapor pipe model, utilizing the code NekRS. Using LES is crucial for comprehending the impact of laminar-to-turbulent transition on pressure recovery.

A simulation framework is created to model the heat pipe's vapor core, laying the foundation for an enhanced understanding of heat pipe behavior. The ultimate goal is to improve and optimize heat pipe designs, provide data to validate lower-fidelity approaches and enhance their performance in nuclear reactors.

## 1. INTRODUCTION

Microreactors are emerging as a promising solution for applications where traditional megawatt-level energy production is impractical. These compact reactors, producing less than 20 MW of power, are designed for autonomous operation in remote locations, military installations, and industrial settings [1]. Of the various microreactor designs, heat-pipe-cooled microreactors, which utilize liquid metal heat pipes, are of particular interest due to their ability to operate at high temperatures and provide passive heat removal [2]. The superior heat transfer capabilities of liquid metal heat pipes allow for more efficient reactor operation compared to traditional heat transfer methods [3].

A heat pipe consists of a sealed tube containing a wick structure attached to the inner surface and a working fluid that continuously transports heat from one end to another by capillary pressure (Figure 1). It comprises three main sections: an evaporator, an adiabatic section, and a condenser. The heat pipe functions by the addition of heat in the evaporator section, causing the fluid to evaporate and flow to the condenser section, where it condenses and releases heat. The fluid then returns to the evaporator by capillary pressure in the wick. The heat pipe's operational range is determined by its heat transport limitations, such as capillary and sonic limits.

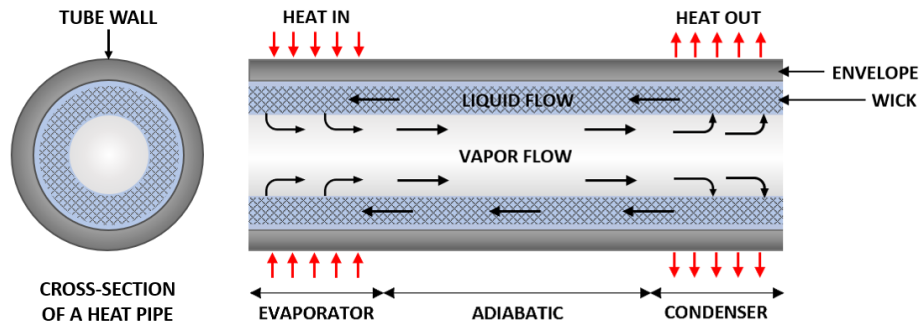

**Figure 1: Heat pipe scheme [4].**

Although liquid metal heat pipe technology is mature [5] its application in the cooling of nuclear reactors is still relatively new, making the development of accurate heat pipe simulation capabilities crucial. A heat pipe code called Sockeye, based on the MOOSE framework, has been developed with the capability to be coupled to other applications and model a microreactor with multiple heat pipes [6]. However, the need for closure models and high-resolution experimental data has been recognized to gain insight into heat pipe behavior and validate lower-fidelity computational approaches.

In annular-wick heat pipes, the flow in the condenser significantly impacts the overall vapor pressure drop, which is crucial for understanding operational limits, particularly the capillary limit. This limit occurs when the applied heat flux causes the liquid in the wick structure to evaporate faster than the capillary pumping power of the wick can replenish it, leading to heat pipe failure.

The capillary force in a heat pipe balances the unrecoverable pressure losses caused by vapor and liquid flow. In annular-wick heat pipes, commonly proposed for nuclear applications, the vapor pressure drop is usually dominant because most of the liquid flows within an annular gap between the wick and the heat pipe wall, rather than in the wick itself [4] [7]. Therefore, accurately predicting the overall vapor pressure drop is essential for understanding heat pipe operational limits.

The importance of studying pressure distribution in heat pipes has been recognized since the early days of heat pipe research [8]. In the evaporator section, the flow accelerates and is mostly laminar, allowing for analytical predictions of the pressure distribution. As the vapor condenses in the condenser section, a partial pressure recovery occurs in the decelerating flow, which transitions to a turbulent regime [9]. Due to the significantly turbulent nature of the flow in the condenser section, empirical correlations are required to predict the pressure distribution. However, experimental data for cylindrical condensers is limited, often outdated, and lacks control of boundary conditions.

This paper aims to develop high-fidelity numerical data of the condenser flow in heat pipes at conditions relevant to alkali heat pipes. The research will focus on better understanding the transition to turbulence in the heat pipe condenser, including its stability and turbulence onset in the vapor flow. As this directly affects pressure recovery, it will contribute towards the development of a mechanistic correlation for vapor pressure drop in the condenser of annular-wick heat pipes.

This study represents a first step in the process with focus on the validation given the limited experimental data available. The vapor flow in the heat pipe is modeled using a simplified porous pipe [10], with mass injection (at the evaporator) and extraction (at the condenser). Two modeling approaches are presented: a two-dimensional RANS model, using Nek5000 [11], and a three-dimensional LES, using the GPU-enabled code NekRS [12]. Both codes are based on the spectral element method, and they have been stablished for nuclear applications [13].

Given the inherent challenges in obtaining a complete understanding of the complex internal physics through experimentation alone, this research will play a crucial role in complementing the knowledge gained from experiments. High-fidelity simulations provide valuable insights into aspects that are difficult to measure or visualize experimentally, such as flow patterns and the effects of various operational parameters on the laminar-turbulent transition in the condenser section.

## 2. METHODOLOGY

The LES and RANS simulations are performed using the spectral-element open-source Computational Fluid Dynamics code Nek5000 [11], and its GPU-enabled version NekRS [12]. The solution is represented as $N^{th}$-order tensor-product polynomials within each element in the quadrilateral (2D) or hexahedral (3D) mesh. These polynomials are constructed on N+1 Gauss-Lobatto-Legendre (GLL) collocation points. Nek5000/RS has demonstrated excellent scalability and has been extensively validated in previous studies [13].

For the LES calculations, an explicit filter is applied to remove energy from the highest wavenumber modes, and the governing equations in the non-dimensional form are given by the Momentum and Continuity Equations:

$$\frac{\partial \mathbf{u}^*}{\partial t^*} + \mathbf{u}^* \cdot \nabla \mathbf{u}^* = -\nabla p^* + \frac{1}{Re} \nabla \cdot \underline{\tau}^* \tag{1}$$

$$\nabla \mathbf{u}^* = \mathbf{0} \tag{2}$$

where and $\underline{\tau}^* = [\mathbf{u}^* \cdot \nabla \mathbf{u}^{*T}]$, $\mathbf{u}^*$ is the non-dimensional velocity, $p^*$ is the non-dimensional pressure, $t$ is time, and $Re$ is the Reynolds number based on the pipe diameter $D$:

$$Re = \frac{\rho u D}{\mu} \tag{3}$$

where $u$ is the inlet velocity, $\rho$ is the density, and $\mu$ is the viscosity.

For the Unsteady RANS (URANS) simulations conducted in this work, the $k-\tau$ turbulence model was used due to its robustness and capability to avoid the numerical discretization errors that could occur with other models, such as $k-\omega$ [14] [15]. The governing equations of this model are given by:

$$\rho \left( \frac{\partial k}{\partial t} + \mathbf{u} \cdot \nabla k \right) = \nabla \cdot (\Gamma_k \nabla k) + P_k - \rho \beta^* \frac{k}{\tau} \tag{4}$$

$$\rho \left( \frac{\partial \tau}{\partial t} + \mathbf{u} \cdot \nabla \tau \right) = \nabla \cdot (\Gamma_\omega \nabla \tau) - \alpha \frac{\tau}{k} P_k + \rho \beta - 2 \frac{\Gamma_\omega}{\tau} (\nabla \tau \cdot \nabla \tau) + C_D \tag{5}$$

$$P_k = \rho \alpha^* k \tau (\underline{S} : \underline{S}), \quad \underline{S} = \frac{1}{2}(\nabla \mathbf{u} + \mathbf{u}^T), \quad C_D = (\delta_d \rho \tau) min \{\nabla k \cdot \nabla \tau, 0\} \tag{6}$$

where $k$ is the turbulent kinetic energy, $\mu_i$ is the diffusion coefficient, $P$ is the production term, $\Gamma_k$ and $\Gamma_\omega$ are diffusion terms, $\beta$ is a coefficient of dissipation, $C_D$ is the cross-diffusion term and $\tau$ is the inverse of the specific dissipation $\omega$, given by:

$$\tau = \frac{1}{\omega} = \frac{k}{\epsilon} \tag{7}$$

with $\epsilon$ being the turbulent dissipation rate.

### 2.1. Model Description

In the vapor core of a heat pipe, the flow is laminar in the evaporator section, but typically transitions to the turbulent regime within the adiabatic or condenser section as it decelerates. This problem of flow transition in heat pipes was generalized by Bowman [10] through an experimental setup using a porous pipe. In Bowman's experiment, air is injected through the porous wall in the first half of the pipe, mimicking the evaporation process in a heat pipe's evaporator section. In the second half of the pipe, air is extracted through the porous wall, simulating the condensation process in a heat pipe's condenser section. This setup allows for a controlled investigation of the flow dynamics and the laminar-to-turbulent transition that occurs within the vapor core of a heat pipe. Figure 2 provides a simplified representation of Bowman's experimental setup, with the porous pipe divided into two sections: the injection section (evaporator) and the extraction section (condenser).

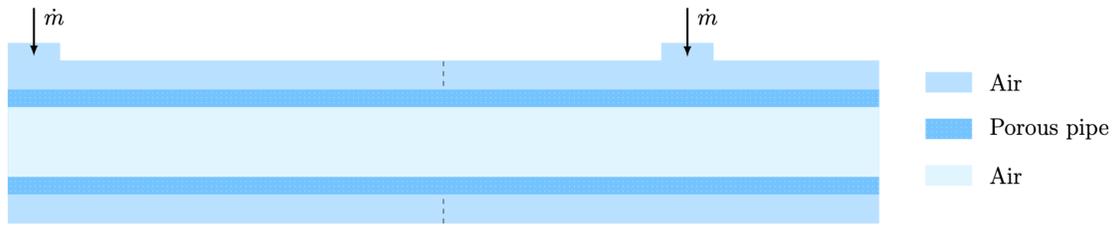

**Figure 2: Scheme of Bowman's vapor pipe experiment.**

#### 2.1.1. RANS Model

A 2D axisymmetric model was developed to simulate the vapor core of the Bowman pipe using RANS simulations. The computational domain was discretized using the *genbox* tool provided by Nek5000. The resulting mesh comprises E = 10,500 elements, with a refinement near the walls. Figure 3(a) illustrates the mesh refinement in the central wall region between the evaporator and condenser sections.

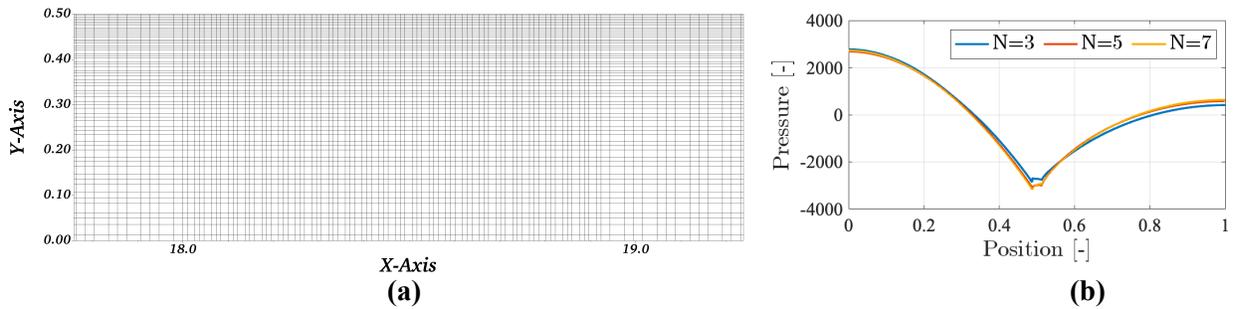

**Figure 3: (a) Cut of the 2D mesh with N=5, showing the refinement near the central wall, and (b) non-dimensional pressure measured in the axial direction at different polynomial orders: N = 3, 5, and 7.**

The inner diameter of the simulated pipe is 0.01651 m, and its total length is 0.6096 m [10]. The simulations were performed using a polynomial order of N = 5, which was determined to be optimal for this study. In fact, a comprehensive mesh convergence study was conducted, as shown in Figure 3(b), and it was determined that the results did not change with a further increase in the polynomial order. Then, the following boundary conditions were incorporated into the model:

- At the evaporator and condenser sections, uniform inlet boundary conditions were prescribed for mass injection and extraction, respectively. For $k$ and $\tau$, Dirichlet boundary conditions were applied at the injection boundary, while Neumann boundary conditions were imposed at the extraction boundary.

- No-slip boundary conditions were enforced at the walls of the adiabatic section and the heat pipe caps. Dirichlet boundary conditions were specified for $k$ and $\tau$ at these locations.

Figure 4 presents a schematic diagram of the computational model used in this study. It is important to note that this model assumes an incompressible flow despite the Mach number being $Ma = 0.17$, where $Ma = u/c$, with $c$ being the speed of sound. This assumption is justified by the relatively low Mach number, which indicates that compressibility effects are not significant in this case. Additionally, the model does not account for temperature or heat flux variations, focusing primarily on the fluid dynamics within the vapor core. The radial Reynolds number, calculated based on the inner diameter of the pipe and the radial velocity, is $Re = 739$. We note that the peak axial Reynolds number is much higher.

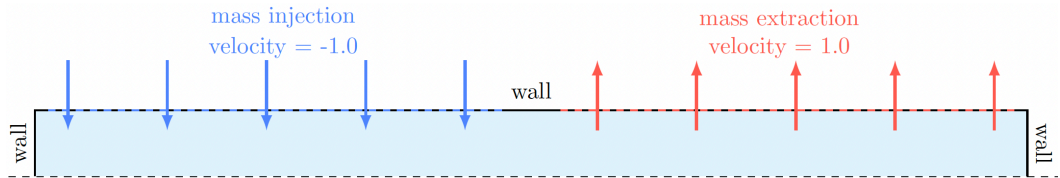

**Figure 4: Scheme of Bowman's vapor pipe RANS model.**

To ensure numerical stability, the computational domain was non-dimensionalized, and unsteady simulations were performed with a time-step on the order of $10^{-5}$. This small time-step size was chosen to maintain a CFL (Courant-Friedrichs-Lewy) number below 0.5, which is a critical criterion for the stability and convergence of the numerical scheme. The simulations were run for one flow-through time, allowing the flow to develop fully within the computational domain. After the completion of the unsteady simulations, the results were time-averaged.

### 2.1.1. LES Model

A 3D mesh was developed using the open-source mesh generator Gmsh [15], composed of 442,860 hexahedral elements, with a refinement near the wall, as shown in Figure 5. The simulations were performed using a polynomial order of $N = 6$, which was optimal for this study, and an explicit filter of 5%. The mesh was designed to solve the Taylor microscale everywhere in the geometry. Once again, the computational domain was non-dimensionalized, and unsteady simulations were performed with a time-step on the order of $10^{-5}$ to maintain the CFL number below 0.5.

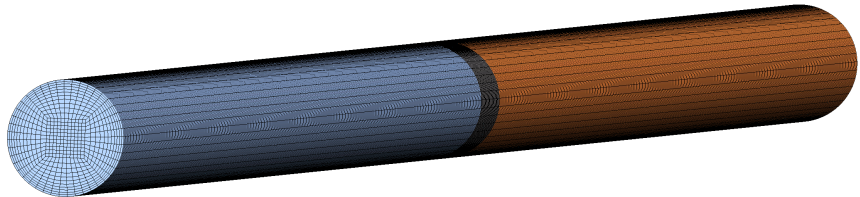

**Figure 5: 3D mesh of Bowman's vapor pipe for the LES model, showing 3 different sections for mass injection, wall, and mass extraction.**

As presented in Figure 5, the following boundary conditions were incorporated into the model:

- At the evaporator and condenser sections, uniform inlet boundary conditions were prescribed for mass injection and extraction, respectively.

- No-slip boundary conditions were enforced at the walls of the adiabatic section and the heat pipe caps.

Figure 6 presents a schematic diagram of the computational model, illustrating the different regions and boundary conditions applied to the domain. To better visualize the various sections of the heat pipe, the regions of the domain in Figure 5 are colored to correspond with the schematic diagram. It is important to note that there is a refinement in the mesh, particularly near the central wall between the evaporator and the condenser sections. This mesh refinement is crucial for accurately resolving the flow dynamics and capturing the laminar-turbulent transition in this region.

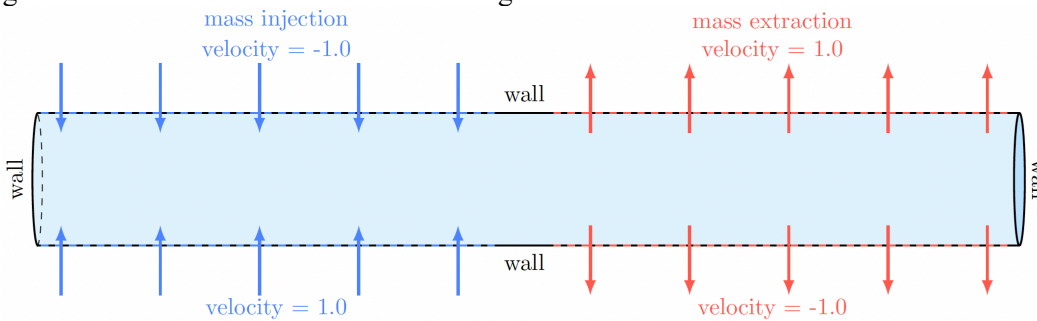

**Figure 6: Scheme of Bowman's vapor pipe LES model.**

### 3. COMPARISONS OF CFD WITH OTHER APPROACHES

Before exploring the results of the CFD analysis in detail, a comparison between CFD and other approaches is provided to put the results in context. In particular, the results from the 2D RANS approach using Nek5000 are compared against two other methods: the analytic solution from Cotter's theory [9] and the Liquid-Conduction Vapor-Flow Model from Sockeye [16]. This Sockeye model solves one-dimensional compressible flow equations for the vapor phase in the heat pipe core and two-dimensional heat conduction equations for the liquid, wick, and cladding.

The simulations were conducted for a horizontal annular-wick heat pipe with sodium as the working fluid. The wick inner diameter was *0.01231* m, and the lengths of the evaporator, adiabatic, and condenser sections were *0.3* m, *0.2* m, and *0.3* m, respectively. Based on the inner wick diameter and radial velocity, the radial Reynolds number was *Re = 312.3*.

For the Nek5000 2D axisymmetric model, a mesh with *6,400* elements was used, with refinement near the wall to account for boundary layer effects. The polynomial order was set to *N = 7*. Figure 7 presents the time-averaged non-dimensional results, which show some key features of the flow:

1. The axial velocity peaks at the middle of the heat pipe and decelerates as it exits the condenser.
2. The turbulent kinetic energy increases as the flow decelerates.
3. As a result of the flow deceleration, the pressure recovers in the condenser section.

These findings provide context for the flow behavior within the horizontal annular heat pipe and the uncertainties associated with the model. They also serve as a basis for a code-to-code comparison of the 2D RANS approach using Nek5000 against other established methods.

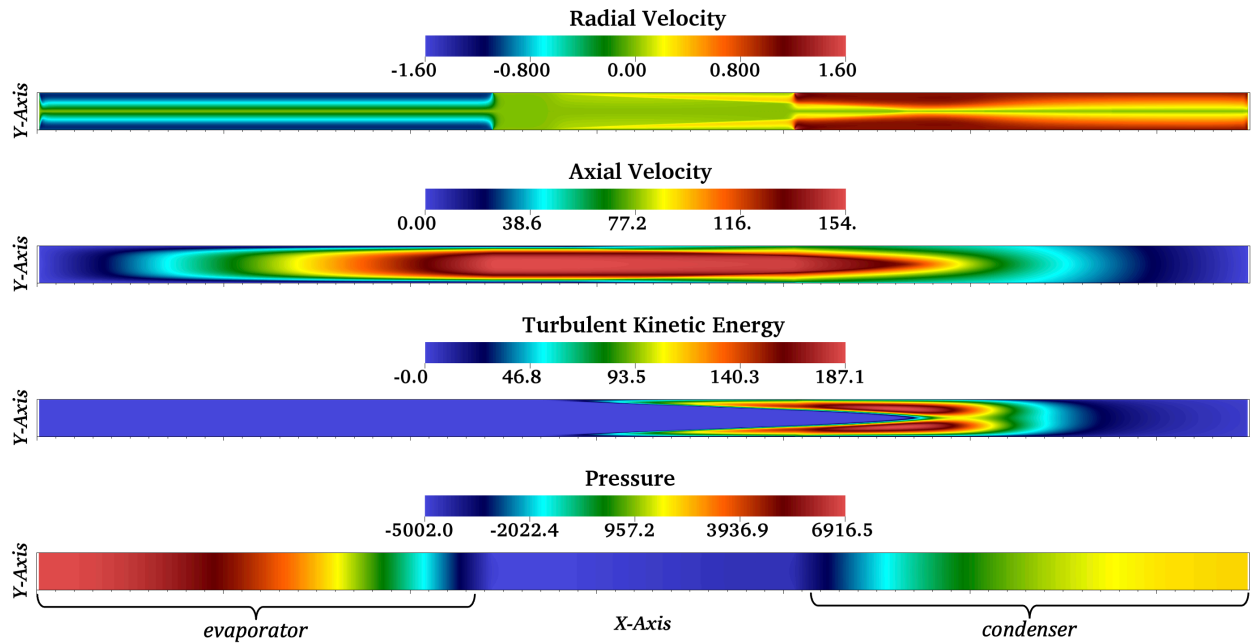

**Figure 7: Non-dimensional results for radial velocity, axial velocity, turbulent kinetic energy, and pressure.**

The Nek5000 simulation results were spatially averaged in the radial direction and converted to dimensional values for comparison with the 1D Analytic and Sockeye results, as presented in Figure 8. Figure 8(b) demonstrates that the Nek5000 simulations perfectly agree with the mass flow rates obtained from the Analytic and Sockeye methods. Furthermore, as shown in Figure 8(a), the Nek5000 simulations accurately reproduced the pressure drop in the evaporator, as predicted by the Analytic method. The Nek5000 results also provided a reasonable estimation of the pressure recovery compared to the other two methods. It is important to note that the analytic method offers only an approximate solution in the condenser region due to the limitations of existing correlations.

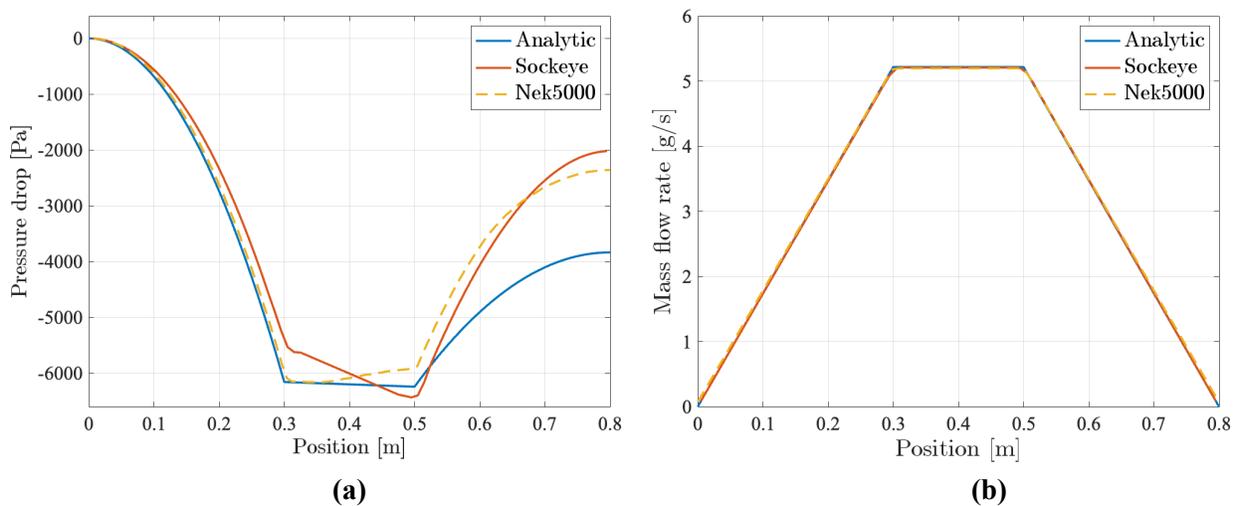

(a)                  (b)

**Figure 8: Comparison of computed and analytic vapor pressure distributions (a) and vapor mass flow rate distributions (b) for the annular-wick heat pipe.**

## 4. DETAILED CFD RESULTS FOR THE BOWMAN EXPERIMENT

Figure 9 presents the time-averaged results for the non-dimensional velocity components, turbulent kinetic energy, and pressure within the computational domain. The radial velocity plot clearly shows the uniform mass injection and extraction boundary conditions. Negative radial velocity values indicate mass injection at the evaporator section, while positive values represent mass extraction at the condenser section.

The axial velocity and turbulent kinetic energy plots reveal a distinct flow pattern along the length of the heat pipe. In the first half of the domain, where mass is injected, the flow experiences a significant acceleration, as evidenced by the increase in axial velocity. As the flow progresses to the second half of the pipe, where mass is extracted, a notable deceleration occurs, causing the axial velocity to decrease accordingly. The turbulent kinetic energy follows a similar trend, increasing in the first half of the domain and dissipating in the second half as the flow decelerates.

The pressure plot showcases a typical feature of heat pipe flow in the condenser: pressure recovery in the second half of the pipe, where the flow decelerates. This pressure recovery phenomenon results from the mass extraction and deceleration processes occurring in the condenser section. The increase in pressure towards the end of the heat pipe is a distinctive characteristic of this type of flow and plays a crucial role in its overall performance and efficiency.

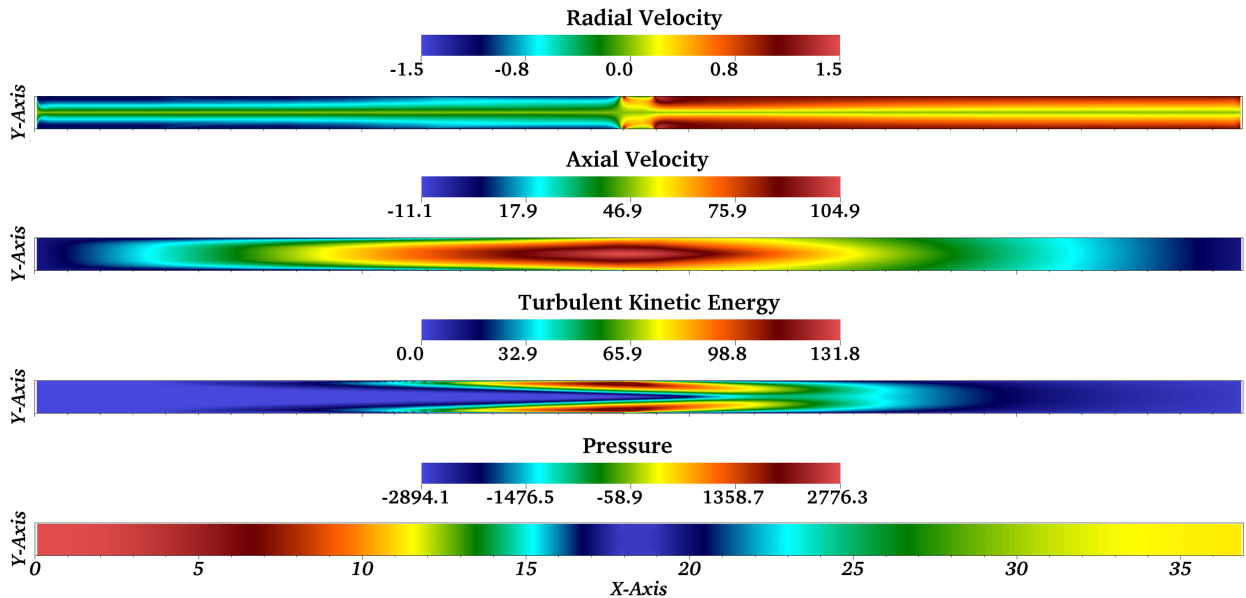

**Figure 9: Non-dimensional results for radial velocity, axial velocity, turbulent kinetic energy, and pressure with scaled by 2 times in the Y-Axis.**

Then, the simulation results were dimensionalized to be compared with the experiments. The velocity profiles along the radial direction were analyzed at two axial locations and compared against experimental data, as shown in Figure 10. The two locations chosen for comparison are at the beginning and the middle of the condenser section. The RANS model agrees well with the experimental results, successfully capturing the velocity profiles at both locations. This comparison validates the ability of the RANS model to accurately represent the flow dynamics within the heat pipe's vapor core, particularly in the condenser section where the laminar-turbulent transition occurs.

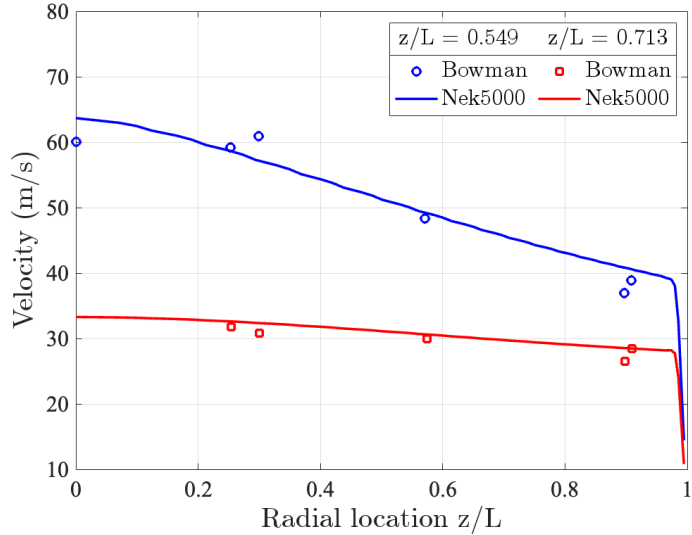

**Figure 10: Velocity as function of axial position.**

Following the validation of the RANS simulations in the 2D domain, the simulation of the 3D LES case was initiated to obtain a detailed representation of the physics in the vapor core. This case was run for multiple flow-through times, allowing the flow to develop fully within the domain. To ensure that the turbulence would not dissipate due to the abrupt mass extraction at the condenser and to provide a better representation of the Bowman experiment, a thin, porous layer was added to the outer surface of the cylinder.

The LES results were then spatially averaged in the radial direction and compared against the RANS and the experimental results. Figure 10 presents the pressure ratio as a function of the axial position within the pipe. Considering the uncertainties associated with the experiment and the manual data extraction from the plot image, the results demonstrate a reasonable agreement between the simulations and the experimental data. The RANS model implemented in Nek5000, employing the $k - \tau$ turbulence model, successfully captured the pressure recovery phenomenon observed in heat pipes operating at a low Mach number. This agreement validates the ability of the chosen modeling approach to accurately represent the flow dynamics and pressure variations within the heat pipe's vapor core. The LES results further validate the modeling approach, showing a better agreement with the experiment, particularly for the condenser region.

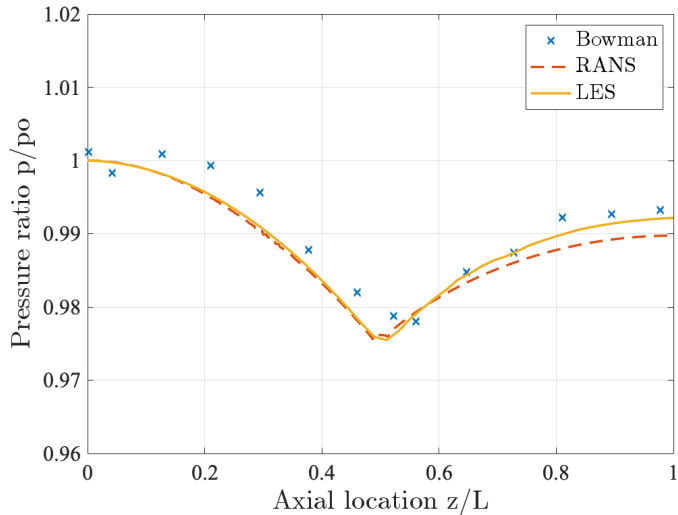

**Figure 10: Pressure ratio as function of axial position.**

Figure 12 illustrates the unique flow pattern within the vapor pipe, resulting from the interplay between mass injection, acceleration, mass extraction, and deceleration. The instant non-dimensional LES results clearly show the transition from laminar flow in the evaporator section to turbulent flow as the fluid exits the condenser. This transition coincides with the beginning of pressure recovery, which begins at the exact location of the turbulence onset. The zoomed-in view of the non-dimensional axial velocity in Figure 12 provides a detailed visualization of this turbulence onset, highlighting the complex flow dynamics within the vapor core.

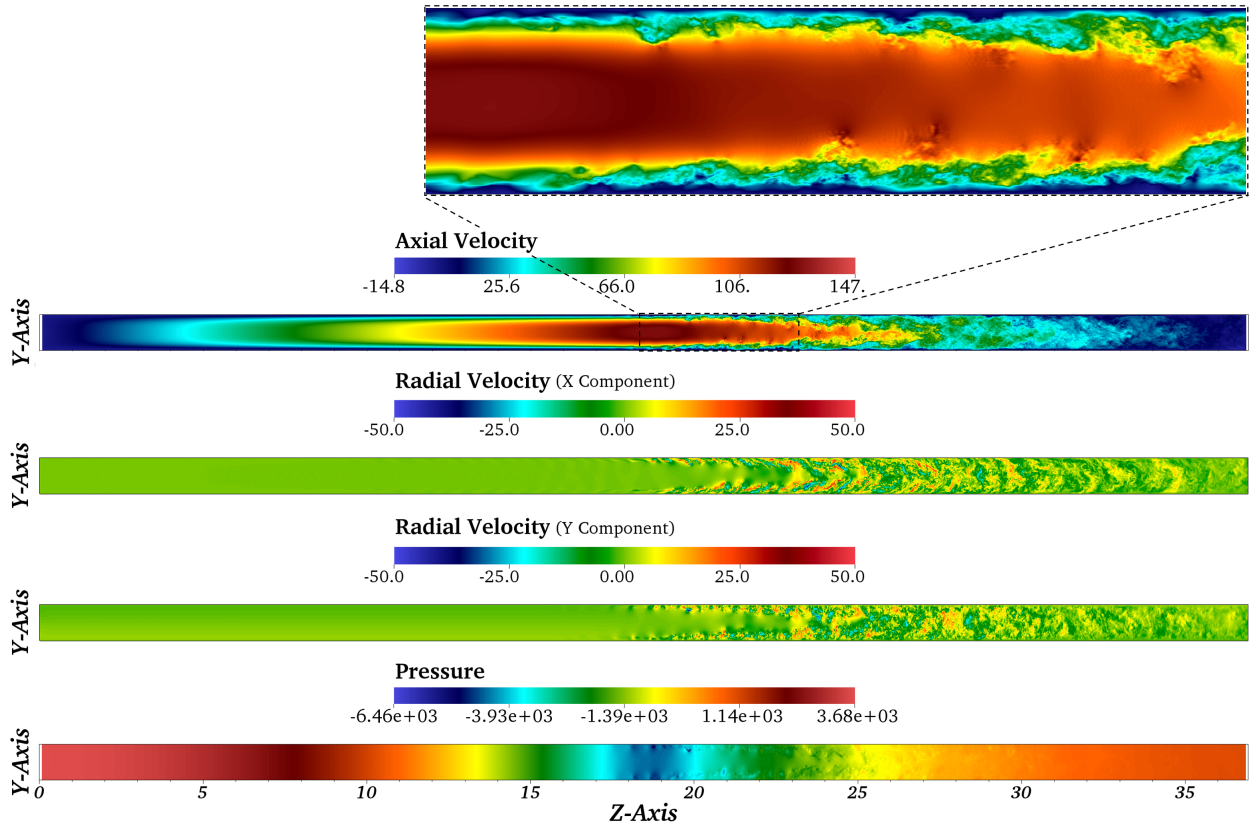

**Figure 12: Instant non-dimensional LES results for the velocity components and the pressure in a cross section of the domain.**

After the convergence of the results, the simulation was time-averaged to provide more detailed insights into the physics of the laminar-turbulent transition in heat pipes. The simulations were time-averaged for ten convective units, and the preliminary non-dimensional results are depicted in Figure 13. Although more averaging time is necessary, as evidenced by the radial velocity profile, these results allow us to make a few observations.

The Turbulent Kinetic Energy (TKE) peak in the LES simulation exhibits a similar trend to that presented for the RANS simulations in Figure 9, but with a magnitude 2 to 3 times higher. Moreover, there is a notable difference in the location of the maximum TKE between the two approaches. In the RANS case, the maximum TKE occurs at the beginning of the condenser section. At the same time, in the LES, this location corresponds to the onset of turbulence, with the TKE peak occurring further downstream in the axial direction. This showcases that the high-fidelity modeling of heat pipes has a potential to improve the RANS models and produce a better representation of the physics involved.

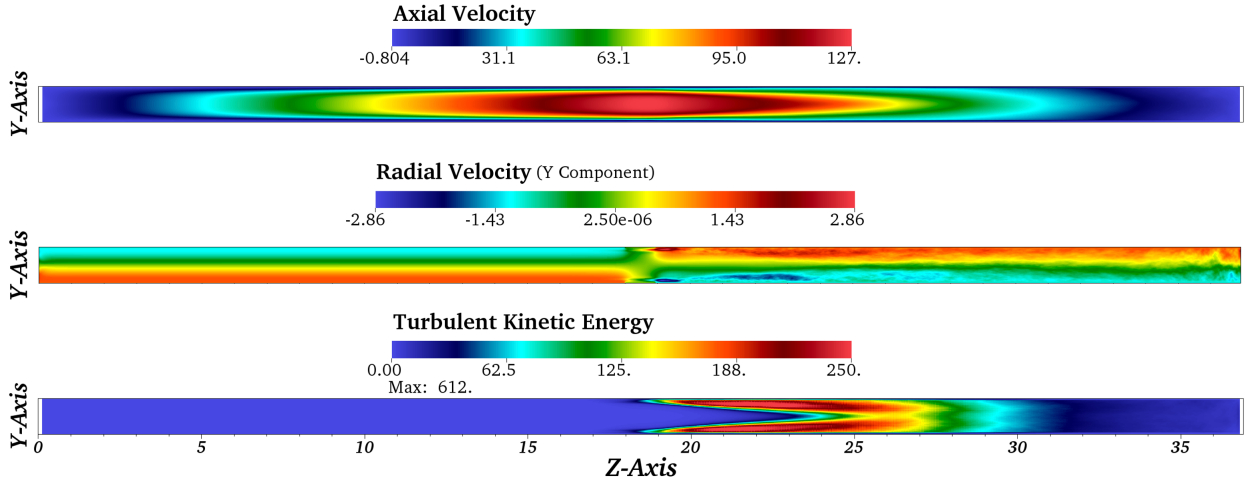

**Figure 13: Time-averaged non-dimensional LES results for the velocity components and the turbulent kinetic energy in a cross section of the domain.**

## 5. CONCLUSIONS

This work aims to explore physics of the vapor core of heat pipes through comprehensive 2D and 3D simulations, with a primary focus on modeling the pressure recovery observed in the condenser section. The ultimate goal is to establish improved correlations for 1D heat pipe codes, enhancing their accuracy and reliability. This manuscript focuses on the validation of RANS simulations, employed to assess the capability of modeling the vapor core using open-source tools and low-cost computational resources while validating the developed methodology against experimental data. Large Eddy Simulations (LES) are then used to capture turbulent flow features in a 3D vapor pipe model, utilizing the code NekRS and following a similar methodology to RANS.

The simulation framework created in this study lays the foundation for an enhanced understanding of heat pipe behavior. The validated numerical results from the simulations using the RANS $k-\tau$ model demonstrate the model's ability to predict the velocity and the pressure recovery phenomena in heat pipes with low Mach numbers, showing reasonable agreement with experimental data. The preliminary 3D time-averaged LES results provide valuable insights into the complex fluid dynamics occurring within the vapor core, including information on the location of the turbulent onset. A more detailed comparative analysis between the RANS and LES approaches will be conducted. The results will also be extended to Direct Numerical Simulations for a more comprehensive evaluation.

The aim of this study is to expand the high-fidelity simulation dataset beyond the limits of the available experimental data, covering a more comprehensive range of parameters. This expanded dataset will provide insights into heat pipe behavior under various conditions, enabling the identification of performance trends and limitations. Furthermore, the Sockeye heat pipe modeling code will be employed to perform validation studies that match numerical setups and available numerical data. By collecting relevant data and analyzing non-dimensional groups, we seek to develop a better mechanistic understanding of the laminar-turbulent transition and its impact on pressure recovery. This improved understanding will support the development of enhanced correlations for unrecoverable pressure losses in heat pipes, ultimately leading to more accurate and reliable predictive tools for heat pipe design and analysis.